\documentclass[aps,pre,reprint,floatfix,superscriptaddress,amsfonts,amsmath,amssymb,citeautoscript]{revtex4-2}
\usepackage[pdftex]{graphicx}

\begin{document}
\title{Phenomenological model of crack patterns in thin colloidal films undergoing desiccation}

\author{Yuri Yu. Tarasevich}
\email[Corresponding author: ]{tarasevich@asu-edu.ru}
\affiliation{Laboratory of Mathematical Modeling, Astrakhan Tatishchev State University, Astrakhan, Russia}

\author{Andrei V. Eserkepov}
\email{dantealigjery49@gmail.com}
\affiliation{Laboratory of Mathematical Modeling, Astrakhan Tatishchev State University, Astrakhan, Russia}

\author{Irina V. Vodolazskaya}
\email{vodolazskaya\_agu@mail.ru}
\affiliation{Laboratory of Mathematical Modeling, Astrakhan Tatishchev State University, Astrakhan, Russia}

\author{Avik P. Chatterjee}
\email{achatter@esf.edu}
\affiliation{Department of Chemistry, SUNY-ESF, One Forestry Drive, Syracuse, New York 13210, USA}
\affiliation{The Michael M. Szwarc Polymer Research Institute, Syracuse, New York 13210, USA}

\date{\today}

\begin{abstract}
A number of geometric and topological properties of samples of crack-template based conductive films are examined to assess the degree to which Voronoi diagrams can successfully model structure and conductivity in such networks. Our analysis suggests that although Poisson--Voronoi diagrams are only partially successful in modeling structural features of real-world crack patterns formed in films undergoing desiccation, such diagrams can nevertheless be useful in situations where topological characteristics are more important than geometric ones. A phenomenological model is proposed that is more accurate at capturing features of the real-world crack patterns.

\end{abstract}

\maketitle

\section{Introduction\label{sec:intro}}

Seamless random metallic networks produced using crack templates are a very promising basis for designing
transparent conductive films (TCFs) since inhomogeneity, dead ends and hot spots are less likely as compared to nanowire-base TCFs~\cite{Chen2024,Voronin2023}. Reliable, stable, and inexpensive technologies for producing such TCFs are well established~\cite{Voronin2023}. Voronoi diagrams (also known as Thiessen polygons) have been used in order to model such crack patterns~\cite{Zeng2020,Kim2022,Tarasevich2023,Esteki2023,Tarasevich2023a,Qiang2024}. In particular, the effect of network regularity upon electro-thermal and optical characteristics has been investigated~\cite{Qiang2024}. Samples of crack patterns have been tested to verify whether or not natural systems can be successfully described by Voronoi diagrams~\cite{Haque2023}. In this context, it is worth pointing out that, Thiessen-polygon metal meshes can be directly  fabricated through nano-imprinting technology~\cite{Song2023}. A Voronoi tessellation is a partition of a space into regions chosen such that the points comprising each region are the nearest (in terms of Euclidean distance) to each of a prescribed set of objects, which we refer to as `seeds'. (see, e.g.,~\cite{Okabe2000}).

Alternatively, desiccation crack patterns can be simulated based on simple physical ideas (so-called a spring network model)~\cite{Richardi2010,Khatun2012,Haque2023,Noguchi2024}. So crack patterns in desiccating colloidal films have been simulated using a spring network model~\cite{Sadhukhan2019}. Recent research suggests that in order to model the crack patterns most accurately, different choices of model (such as Gilbert or Voronoi tessellations or iterative cell division) are appropriate for patterns that arise in different types of material~\cite{Roy2022}.

In the case of TCFs, the main quantities of interest are the transparency and the sheet resistance.
Although the \emph{exact} electrical resistance of a network can be directly calculated if the network structure and the resistance of each wire segment are known, analytical results for the dependence of the resistance on physical and geometrical parameters are difficult to extract. By contrast, analytical approaches make it possible to obtain the \emph{average} electrical resistance of a class of networks if the topology of networks and the distribution of the wire segment resistances are known.

\citet{Kumar2016} proposed a formula describing the dependence of the sheet resistance of a two-dimensional (2d) random resistor network (RRN) on the main physical parameters, viz,
\begin{equation}\label{eq:Kumar}
  R_\Box = \frac{\pi\rho}{2 A \sqrt{n_\text{E}}},
\end{equation}
where $\rho$ is the resistivity of the material, $A$ is the cross-section of the wire, and $n_\text{E}$ is the number of wire segments per unit area. Although the authors claimed the method as purely geometrical, in fact, their approach employs a mean-field approximation (MFA). In the case of RRNs, the MFA deals with only a single conductor placed in the mean electric field that all other conductors produce, instead of explicit consideration of each of the conductors in the system.

The approach is based on the fact that in a dense, homogeneous and isotropic two-dimensional RRN, the electric potential varies approximately linearly between two electrodes when a potential difference is applied to opposite boundaries of such an RRN~\cite{Sannicolo2018}. \citet{Kumar2016} supposed that when a potential difference is applied across the opposite borders of a sample, the number of conductive wires intersecting an equipotential line is $\sqrt{n_E}$ per unit length. This assumption overestimates the intersection number by a factor of $\approx 1.5$~\cite{Tarasevich2023}. A more accurate result for the sheet resistance based on a MFA
\begin{equation}\label{eq:Rsheet}
R_\Box = \frac{2 \rho}{ n_\text{E}  \langle l \rangle A}
\end{equation}
additionally includes the mean length of conductive wires, $ \langle l \rangle$~\cite{Tarasevich2019}.
In fact, $\langle l \rangle = C n_E^{-1/2}$, where the value of the constant $C$ depends on the particular kind of the network, e.g., $C = \sqrt{\pi/2}$ for dense random nanowire networks, $C = \sqrt{2}$ for a square lattice, $C = \sqrt{2/\sqrt{3}}$ for a honeycomb lattice, and, in the case of Poisson--Voronoi diagrams (PVDs), it equals the edge length expectation at $n_E = 1$, $C \approx 1.154$.
Although \eqref{eq:Rsheet} is in closer agreement with the results of direct calculations of electrical conductance than~\eqref{eq:Kumar}, both formulas overestimate the electrical conductivity. Each of these approaches~\cite{Kumar2016,Tarasevich2019} treated the RRNs under consideration as being isotropic and on average homogeneous. However, in real systems, local fluctuations of the number of edges per unit area are unavoidable. Unfortunately, accounting for these fluctuations leads to only an insignificant improvement of the estimate for the electrical conductivity~\cite{Tarasevich2023a}. Possibilities for improving estimates of the  electrical conductivity using the MFA seem to us to be completely exhausted.

Alternatively, the effective medium theory (EMT)~\cite{Bruggeman1935} is often applied to predict physical properties, e.g., electrical conductance, of disordered systems including RRNs~\cite{Kirkpatrick1973,Clerc1990,Luck1991,Sahimi1997,OCallaghan2016,He2018,Zeng2022} (a systematic description can be found in Ref.~\onlinecite{Choy2015}).
The main ideas of applying the EMT to networks with regular structures (such as square, honeycomb, and triangular lattices) and random conductances of edges (0 or 1) have been presented in the works~\cite{Kirkpatrick1971,Kirkpatrick1973,Joy1978,Joy1979,Clerc1990}. By contrast, a square lattice of resistors where conductance of the resistors corresponded to the truncated Gaussian distribution has been studied using the EMT and computer simulation~\cite{Melnikov2018}. When the Gaussian distribution is characterized by the mean value $g_0$ and the standard deviation $0.2g_0$, the effective resistance is $R_\text{eff} = 1.021 g_0^{-1}$.

In each of the above-mentioned cases, assumptions of symmetry and homogeneity of the networks played an important role in the approximate treatments. Alternatively, a more formal and general consideration based on Foster's theorem~\cite{Foster1961} is possible~\cite{Marchant1979}. For a regular network of valence $z$ (a $z$-regular network) with different branch admittances (complex conductances) $y$, the following approximation is valid~\cite{Marchant1979}
\begin{equation}\label{eq:MarchantEMT}
  \left\langle \frac{y_m - y}{y + y_m (z/2 -1)} \right\rangle \approx 0,
\end{equation}
where the brackets $\langle\cdot\rangle$ denotes an average over the distinct values of individual resistors in the network. The application of EMT to regular networks is equivalent to replacement of the distribution of $y$, by a unique value $y$ obtained in an `effective network' of the \emph{same structure}, filled with identical conductances $y_m$ and given by~\eqref{eq:MarchantEMT}. However, the application of EMT to regular networks introduces an error that grows with the broadening of the real distribution of branch admittances, viz., the broader this distribution the larger the error.

The present study investigates of the morphology of the crack-template-based (CTB) TCFs in order to assess which model for crack patterns is best applicable to this particular sort of crack patterns. Using an appropriate model, artificial networks can be generated to mimic the properties of real-world CTB TCFs. An EMT can then be used to obtain a dependence of the electrical conductance of networks under consideration on the most important physical parameters characterizing the network.

The rest of the paper is constructed as follows. Section~\ref{sec:methods} describes some known topological and geometrical properties of desiccation crack patterns as well as technical details of our simulation. Section~\ref{sec:results} presents the analytical approach, together with our main findings. Section~\ref{sec:concl} summarizes the main results.

\section{Background and Methods\label{sec:methods}}
\subsection{Some topological and geometrical properties of desiccation crack patterns}
Results of image processing of published photos of CTB TCFs reveals that in these crack patterns the  overwhelming majority of nodes are of valence equal to 3~\cite{Tarasevich2023}. The small fraction of nodes that have valence equal to 1 corresponds to dead ends, representing the termini of edges that do not contribute to the electrical conductivity. In addition to dead ends, boundaries of photos also produce apparent nodes with valence 1. Bends in wires correspond to the small fraction of nodes with a valence equal to 2. Nodes with an apparent valence greater than 3 (also a small fraction) should be treated as an artifact of image processing of photos with  modest resolution when two or more nodes that are very near each other appear as only one node, since simple mechanical arguments suggest that X-shaped cracks are unlikely. Thus, to a first approximation, networks with valence 3 (3-regular networks) are an appropriate tool for modeling CTB TCFs.

An instance of a random 3-regular network is a PVD (random plane Voronoi tessellation). In this particular case, points (seeds) are randomly distributed within a bounded domain on a plane. In a plane PVD,  the valence of each vertex is~3, while the average number of vertices in a cell is 6~\cite{Meijering1953}. Thus, in the terminology of the graph theory, a random plane Voronoi tessellation generates a 3-regular planar graph.

\begin{figure}[!htb]
  \centering
  \includegraphics[width=\columnwidth]{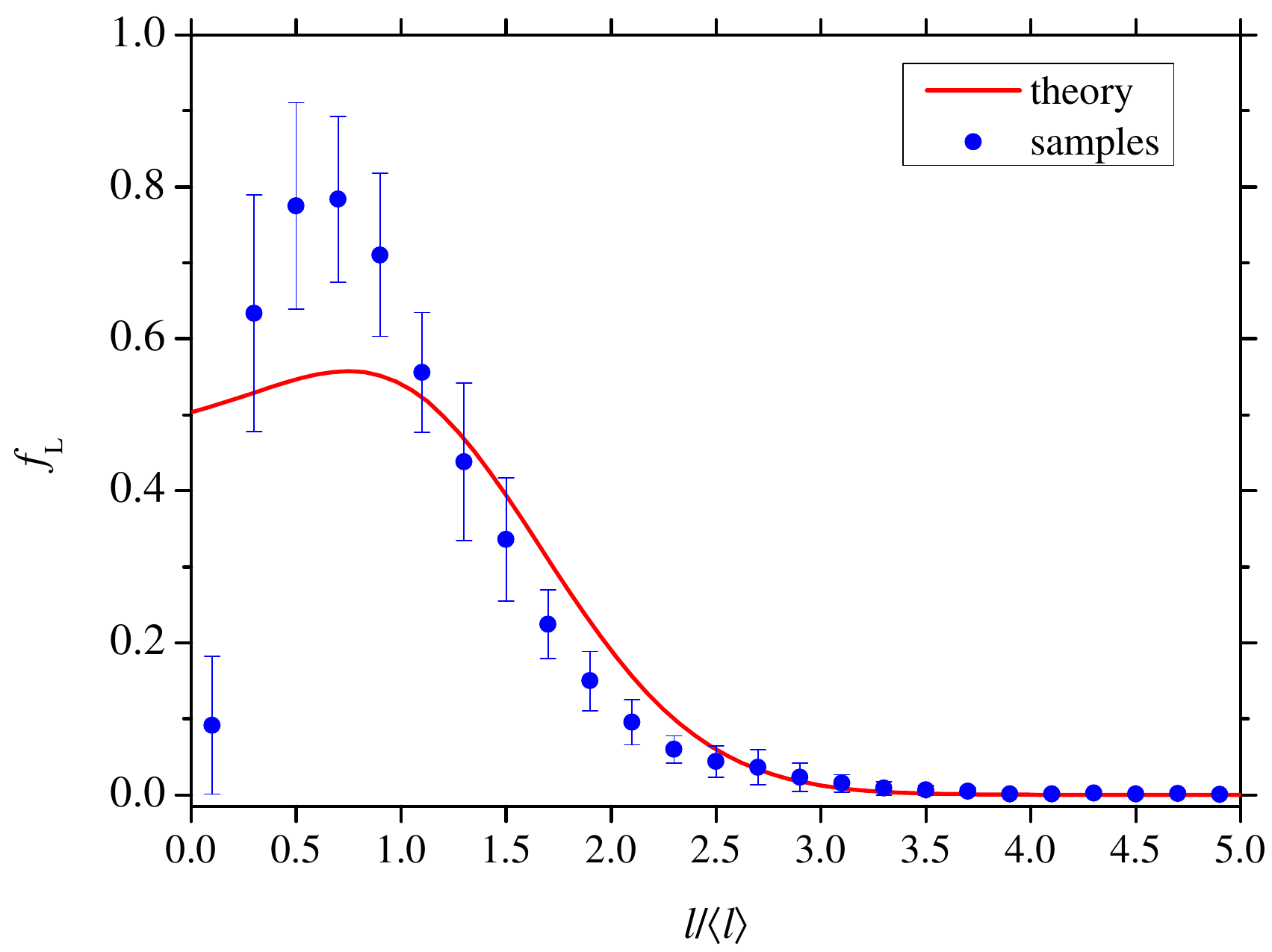}
  \caption{PDF for the lengths of PVD edges~\cite{Brakke2005} (line) along with corresponding densities of real-world crack patterns (markers)~\cite{Tarasevich2023}. Error bars correspond to standard error of the mean (SEM).}\label{fig:PDFBrakkeSamples}
\end{figure}

However, geometrical properties of PVDs and real-world crack patterns, viz., angles between edges and length distributions of edges, are  similar but not identical~\cite{Tarasevich2023}. While it is unlikely that the angles between conductive edges has a substantial impact upon the electrical conductivity of networks, the length distribution of edges is an important factor to take into consideration. Figure~\ref{fig:PDFBrakkeSamples} demonstrates the probability density function (PDF) for the lengths of edges of Voronoi diagrams~\cite{Brakke2005} and the corresponding densities of real-world desiccation crack patterns~\cite{Tarasevich2023}. We surmise that the significant deviation between the PDFs for the shortest edges arises due to shortcomings in the accuracy of the image processing when edges are short.  In fact, accounting for modest resolution of images, short edges are difficult to accurately identify, enumerate, or measure.

Given that the topological and geometrical properties of PVDs and crack patterns are similar although not identical, PVDs seem to be a reasonable starting point for modeling crack patterns when topological rather than geometric quantities are of the greatest importance~\cite{Zeng2020,Kim2022,Tarasevich2023,Esteki2023,Tarasevich2023a}. However, when quantitative estimates for geometrical properties such as the edge length distribution are of importance, it is not inevitable that PVDs will invariably best describe the crack pattern morphology.

\subsection{Sampling and computations}
For analysis, we choose 5 images of real CTB networks. We refer to
the real-world networks as sample 3, sample 4, and sample 5~\footnote{The unpublished photos were kindly provided by A.S.~Voronin.}; sample 1 corresponds to \cite[Fig.5d]{Gupta2014}; and sample 2 corresponds to \cite[Fig.2a]{Rao2014}. Table~\ref{tab:samples} presents the main geometrical parameters that characterize the samples.
\begin{table}[!htb]
  \centering
  \caption{Characterization of samples. $L_x$ and $L_y$ correspond to the domain size,  $N_E$ is the number of edges, $N_s$ is the number of faces, $\langle l \rangle$  and $\sigma_l$ are the mean edge length and its standard deviation. All lengths are indicated in pixels.\label{tab:samples}}
\begin{ruledtabular}
\begin{tabular}{lcccccc}
    & $L_x$ & $L_y$ & $N_E$ & $N_s$ & $\langle l \rangle$ & $ \sigma_l$\\
  \hline
sample 1 & 1022 & 754 & 890  & 329 & 33 & 18\\
sample 2 & 1023 & 953 & 1936 & 706 & 25 & 14\\
sample 3 & 1023 & 719 & 2232 & 824 & 21 & 12\\
sample 4 & 1023 & 719 & 834  & 314 & 36 & 23\\
sample 5 & 1023 & 719 & 1524 & 564 & 25 & 13\\
\end{tabular}
\end{ruledtabular}
\end{table}
All samples are fairly isotropic, since the nematic order parameter is of order of 0.01 for any sample.

\begin{figure}[!htb]
  \centering
  \includegraphics[width=0.9\columnwidth]{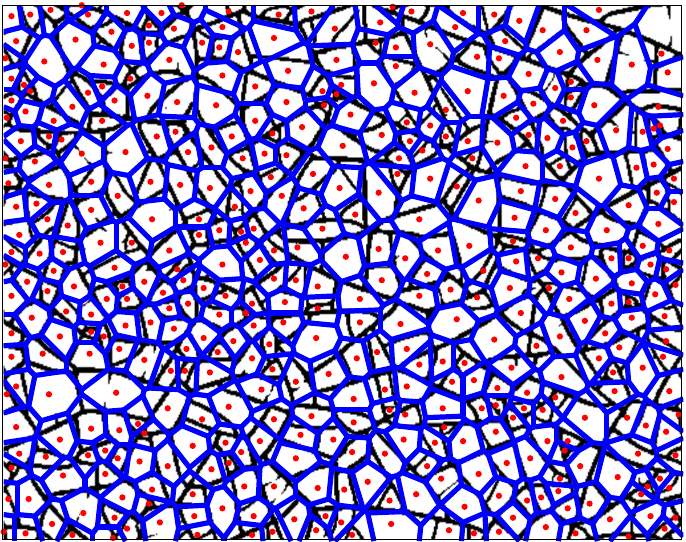}
  \caption{A preprocessed image of the particular crack template (the original photo was kindly provided by A.S.Voronin) along with its AVD.}\label{fig:CracksVoronoi}
\end{figure}

Figure~\ref{fig:CracksVoronoi} demonstrates a particular crack pattern along with a corresponding Voronoi diagram derived from this pattern~\footnote{See Fig. S1 in Supplemental Material for details of image processing.}. To process the photo of the crack pattern, we used StructuralGT~\cite{Vecchio2021}, slightly modified to match our particular requirements. The network so obtained was refined, viz., dead-ends were removed, and pairs of edges meeting in vertices of valence 2 were combined into a single straight edge. After that, all faces (cells) were identified and the locations of their centroids were calculated. These centroids were used as seeds to generate a Voronoi diagram. It should be noted that these seeds are not, however, centroids of the faces of the Voronoi diagram constructed in this fashion. Hereinafter we will denote such diagrams as accompanying Voronoi diagrams (AVDs). This method is based on the approach described in Ref.~\onlinecite{Roy2022}. Fairly visible difference between the initial crack template and its AVD suggests that this particular crack template cannot be modeled by a centroidal Voronoi diagram (CVD), despite  the fact that CVDs have been  successfully used to model the crack patterns of other origins, such as columnar rock cracks~\cite{Meng2018} and mud/clay cracks~\cite{Haque2023}.

To create RRNs that are based upon underlying Voronoi diagrams, a square domain $L \times L$ with periodic boundary conditions (PBCs) was used. $N_s$ points (seeds) were randomly placed within this domain (torus), then, a Voronoi tessellation was performed. After that, PBCs were removed, i.e., the torus was unfolded into a square. The edges of the obtained Voronoi diagram were treated as resistors. To calculate the effective conductance of this RRN, a potential difference was applied across the opposite boundaries of the domain.

To calculate the electrical conductance of the network, we attached a pair of superconducting buses to the two opposite boundaries of the domain in such a way that the potential difference was applied either along the $x$ or $y$ axes. Applying Ohm’s law to each resistor (edge) and Kirchhoff's point rule to each junction (vertex), a system of linear equations was obtained. This system was solved numerically. For the artificial computer-generated networks, we used $L=1$ and 100 independent samples were generated for each value of the number density of seeds. For each value of the number density of seeds, the effective conductivity was averaged over 100 independent runs and over both directions $x$ and $y$.

Since in the real-world samples $L_x \ne L_y$, the resistances along $x$ and $y$ axes are different. The effective conductivity, $G$, can be calculated as follows.
\begin{equation}\label{eqGeff}
  G = \frac{L_x}{R_x L_y}, \quad G = \frac{L_y}{R_y L_x}.
\end{equation}
When the error bars are not shown explicitly in a plot, the standard error of the mean is of the order of the marker size.

\section{Results\label{sec:results}}
\subsection{Properties of crack pattern networks}
\begin{figure}[!htb]
  \centering
  \includegraphics[width=\columnwidth]{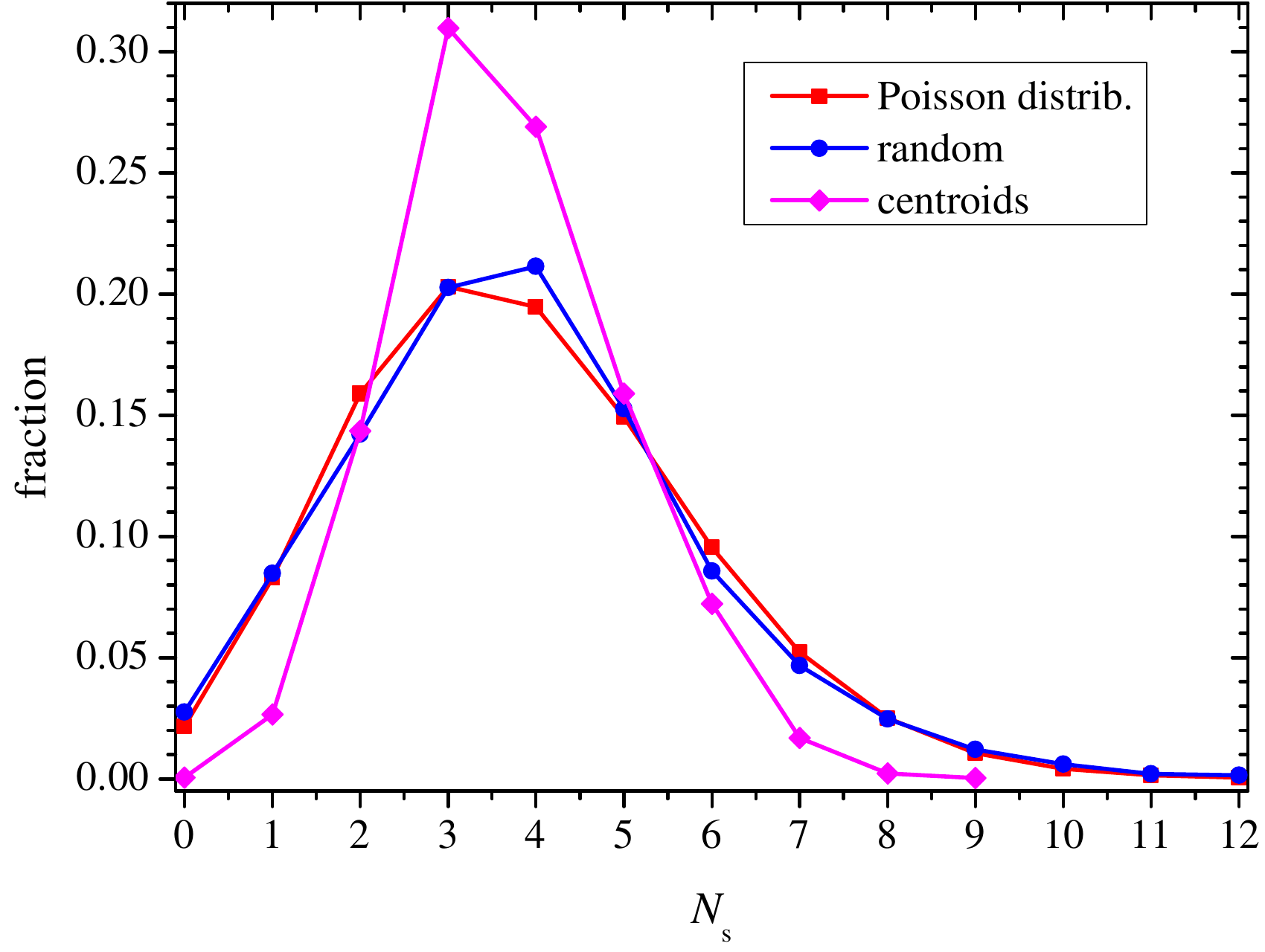}
  \caption{Probability to find the given number of points (centroids of the network faces and randomly placed points), $N_s$, within a randomly placed window along with a corresponding Poisson distribution for an equal number density of points.}\label{fig:TestPoisson}
\end{figure}

Based on the appearance of the samples, we did not analyze their resemblance to patterns generated by either to the Gilbert tessellation or iterative cell division, but focused instead on their similarity to the Voronoi diagrams. Although there are ways to estimate the Voronoi-ness of a particular network~\cite{Haque2023}, our investigation focused upon whether or not the centroid distribution can be considered as being Poisson. To perform this check, we counted the number of centroids, $N_s$, within each of 10\,000 randomly placed rectangular windows each of the same size. Figure~\ref{fig:TestPoisson} demonstrates that the distribution of the number of centroids within a window is narrower as compared to the Poisson distribution. By contrast, if the same number of points is randomly deposited within the same region, the resultant distribution is (as expected) much closer to the Poisson distribution (Fig.~\ref{fig:TestPoisson}).

Using randomly placed windows of different sizes, we found that, for the real-world samples, the variance, $\sigma_N^2$, is 2--3 times smaller than the mean $\langle N_s \rangle$, while for the Poisson distribution $\sigma_N^2 = \langle N_s \rangle$ (Tab.~\ref{tab:Poisson}). Hence, centroids of the faces of the desiccated crack patterns do not obey the Poisson distribution and the AVD obtained from analysis of a real sample is not a PVD. The regularity~\cite{Zhu2001} has also been calculated
\begin{equation}\label{eq:regularity}
  \delta = \frac{r_\text{min}}{s}, \quad s = \sqrt{\frac{2 L_x L_y}{N_s \sqrt{3}}},
\end{equation}
where $r_\text{min}$ is the minimal distance between nearest neighbors (maximal `inhibition distance'), $s$ the distance between seeds in a regular honeycomb lattice with the same number of faces per unit area.
\begin{table}[!htb]
  \centering
  \caption{Characterization of centroid distribution. $\langle N_s \rangle$ is the  mean value for the number of centroids within a randomly placed window, $\sigma_N^2$ is its variance, $r_\text{min}$ is the minimal distance between the nearest neighbors, $\langle r \rangle$ is the mean distance between the nearest neighbors, $\sigma^2_r$ is its variance, $r_\text{max}$ is the maximal distance between the nearest neighbors, $\delta$ is the regularity~\eqref{eq:regularity}.  \label{tab:Poisson}}
\begin{ruledtabular}
\begin{tabular}{lcccccc}
    & $\sigma_N^2/\langle N_s \rangle$ & $r_\text{min}$ & $\langle r \rangle$ & $\sigma^2_r$ & $r_\text{max}$ & $\delta$ \\
   \hline
sample 1 & 0.349(20) & 18 & 39 & 64 & 63 & 0.35 \\
sample 2 & 0.396(13) & 16 & 30 & 40 & 71 & 0.40 \\
sample 3 & 0.438(11) & 8  & 23 & 27 & 41 & 0.25 \\
sample 4 & 0.533(16) & 17 & 37 & 85 & 77 & 0.33 \\
sample 5 & 0.323(10) & 10 & 29 & 39 & 49 & 0.26 \\
\end{tabular}
\end{ruledtabular}
\end{table}

\begin{figure}[!htb]
  \centering
  \includegraphics[width=\columnwidth]{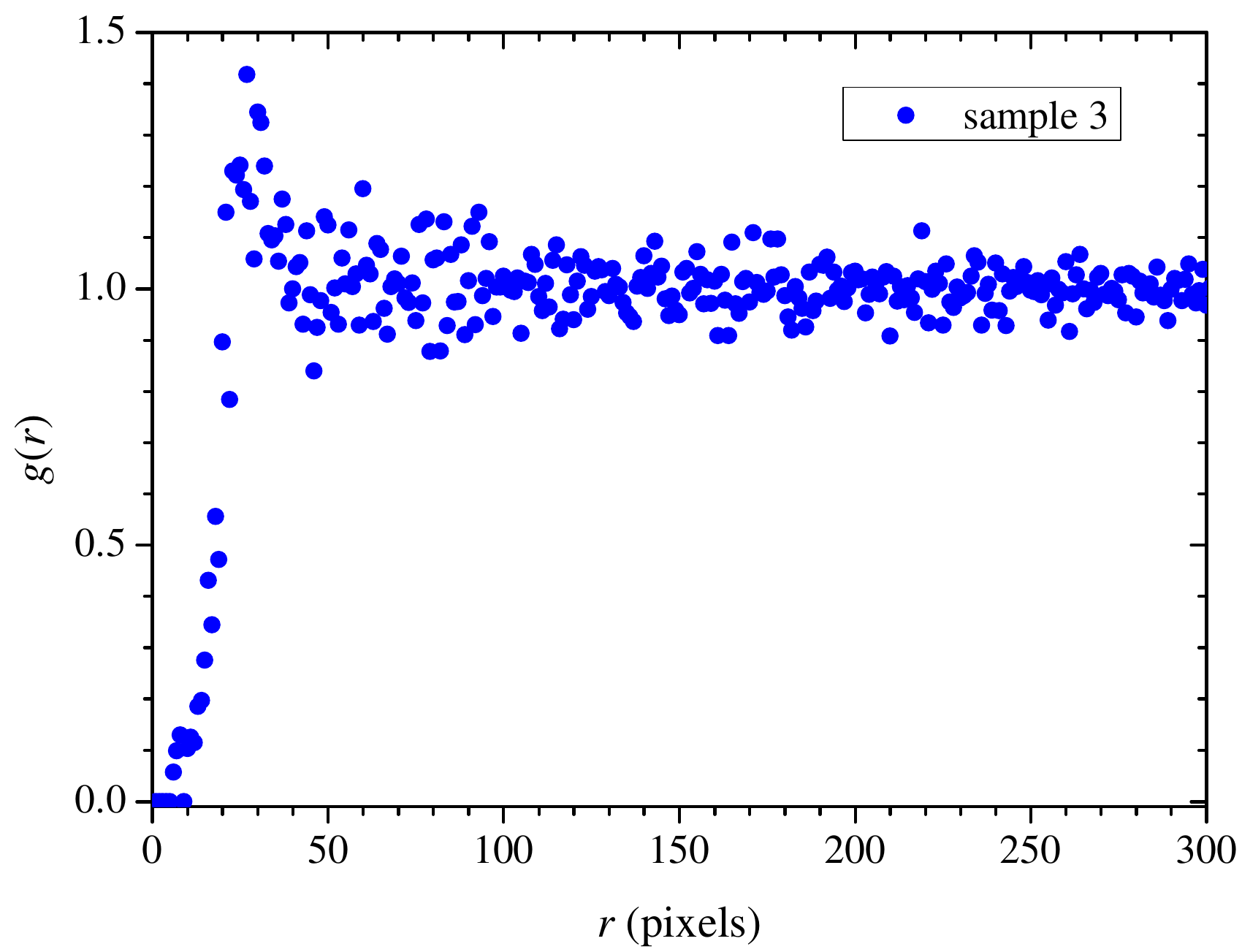}
  \caption{Example of the radial distribution function calculated for a particular crack pattern. }\label{fig:RDF}
\end{figure}

In order to reveal the possible correlations in the positions of the centroids, we calculated the radial distribution function also known as the radial distribution function (RDF). The RDF, $g(r)$, indicates how many centroids are within a distance between $r$ and $ r+dr$ away from a specific, selected, centroid. Figure~\ref{fig:RDF} demonstrates the absence of obvious long-range correlations. There is  a peak at short distance followed by a monotonic decrease with increasing values of $r$.

\begin{figure}[!htb]
  \centering
  \includegraphics[width=\columnwidth]{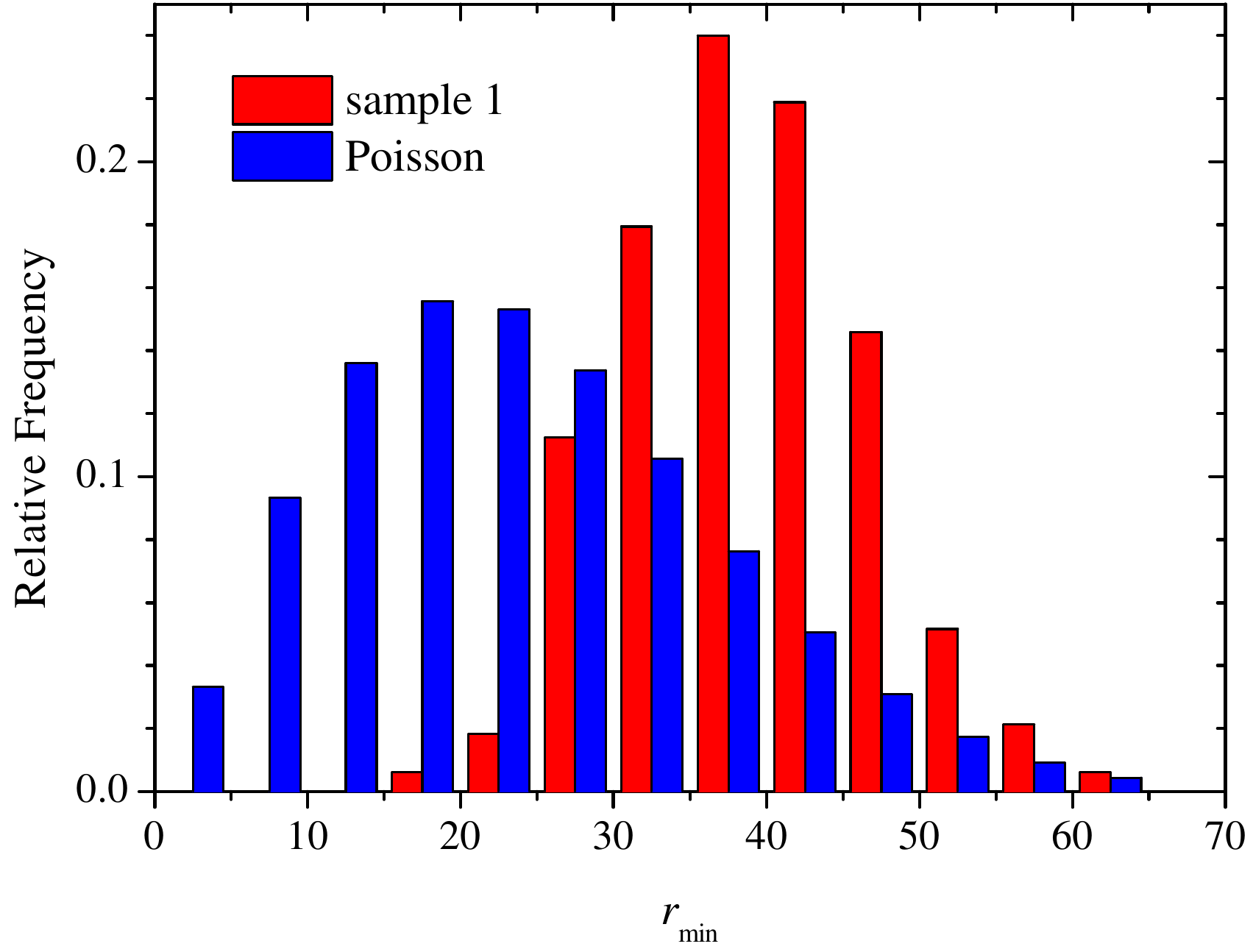}
  \caption{Distribution of the distances between nearest centroids in one particular crack pattern. Corresponding distribution for randomly placed points is presented for  comparison.}\label{fig:mindist}
\end{figure}

In order to better understand the peak at small separations, we next examined the distribution of distances between pairs of nearest neighbor centroids. An example of the distribution is presented in Fig.~\ref{fig:mindist}; detailed information for each sample is presented in Tab.~\ref{tab:Poisson}. The distribution approximately resembles a Gaussian distribution; the mean distance is $31.5 \pm 0.4$, there is a `dead zone' around each centroid, i.e., the distance between centroids cannot be less than 9.86. This distribution is clearly consistent with Fig.~\ref{fig:RDF}. However, this behavior is very different from what is observed in a typical random location, viz., when the RDF is $f(r_\text{min}) = 2\pi r_\text{min} n \exp(-\pi n r_\text{min}^2)$.

`Dead zones' around centroids suggest that the distribution of centroids might be a hyperuniform one. To investigate this possibility, we calculated the number of centroids in a circular observation window with radius $r$ and its variance $ \sigma^2(r)$. If the variance increases more slowly than the area of the observation window that is,
\begin{equation}\label{eq:HU}
  \lim_{r\to \infty} \frac{\sigma^2(r)}{r^2} = 0,
\end{equation}
the distribution is said to be hyperuniform~\cite{Torquato2018}. Due to the finite (and modest) size of the available images of the crack patterns, \eqref{eq:HU} can be checked only approximately. However, our computations demonstrated that the variance scales in the same manner as the radius of the observation window~\footnote{See Fig. S2 in Supplemental Material for the variance, $ \sigma^2(r)$, of number of centroids in a circular observation window against the window radius, $r$.}. Hence, the possibility that the distribution of centroids is hyperuniform is discounted.

Additionally, the structure factor was calculated. The structure factor is closely related to both the RDF and the hyperuniformity~\cite{Hansen2013}. Since neither long-range order nor hyperuniformity have been revealed, the structure factor quite expectedly demonstrated only a random noise.

In order to characterize the initial crack networks and AVDs, we used the circularity~\cite{Richard2001} also known as the isoperimetric quotient. The circularity of one particular shape is
\begin{equation}\label{eq:Circularity}
  Q = \frac{4 \pi\langle  A\rangle}{\langle C^2\rangle},
\end{equation}
where $C$ is the perimeter, $A$ is the area of a face. For regular hexagons $Q = 0.907$. Table~\ref{tab:circularuty} presents the circularity averaged over all faces both for the initial crack networks and the AVDs. Table~\ref{tab:circularuty} demonstrates that the crack templates less resemble a honeycomb network as compared to AVDs.
\begin{table}[!htb]
  \centering
  \caption{Circularity.\label{tab:circularuty}}
\begin{ruledtabular}
\begin{tabular}{lcccccc}
    & $Q$ & $\sigma^2_Q$ & $Q$ & $\sigma^2_Q$  & $Q$ & $\sigma^2_Q$ \\
    & \multicolumn{2}{c}{crack networks}& \multicolumn{2}{c}{AVDs}& \multicolumn{2}{c}{model \eqref{eq:Pspring}}\\
  \hline
sample 1 & 0.74 & 0.11 & 0.80 & 0.07 & 0.767 & 0.006 \\
sample 2 & 0.75 & 0.11 & 0.80 & 0.06 & 0.768 & 0.006 \\
sample 3 & 0.72 & 0.11 & 0.80 & 0.06 & 0.769 & 0.006 \\
sample 4 & 0.65 & 0.12 & 0.79 & 0.07 & 0.764 & 0.006 \\
sample 5 & 0.73 & 0.13 & 0.80 & 0.07 & 0.770 & 0.006 \\
\end{tabular}
\end{ruledtabular}
\end{table}

\begin{figure}[!htb]
  \centering
  \includegraphics[width=\columnwidth]{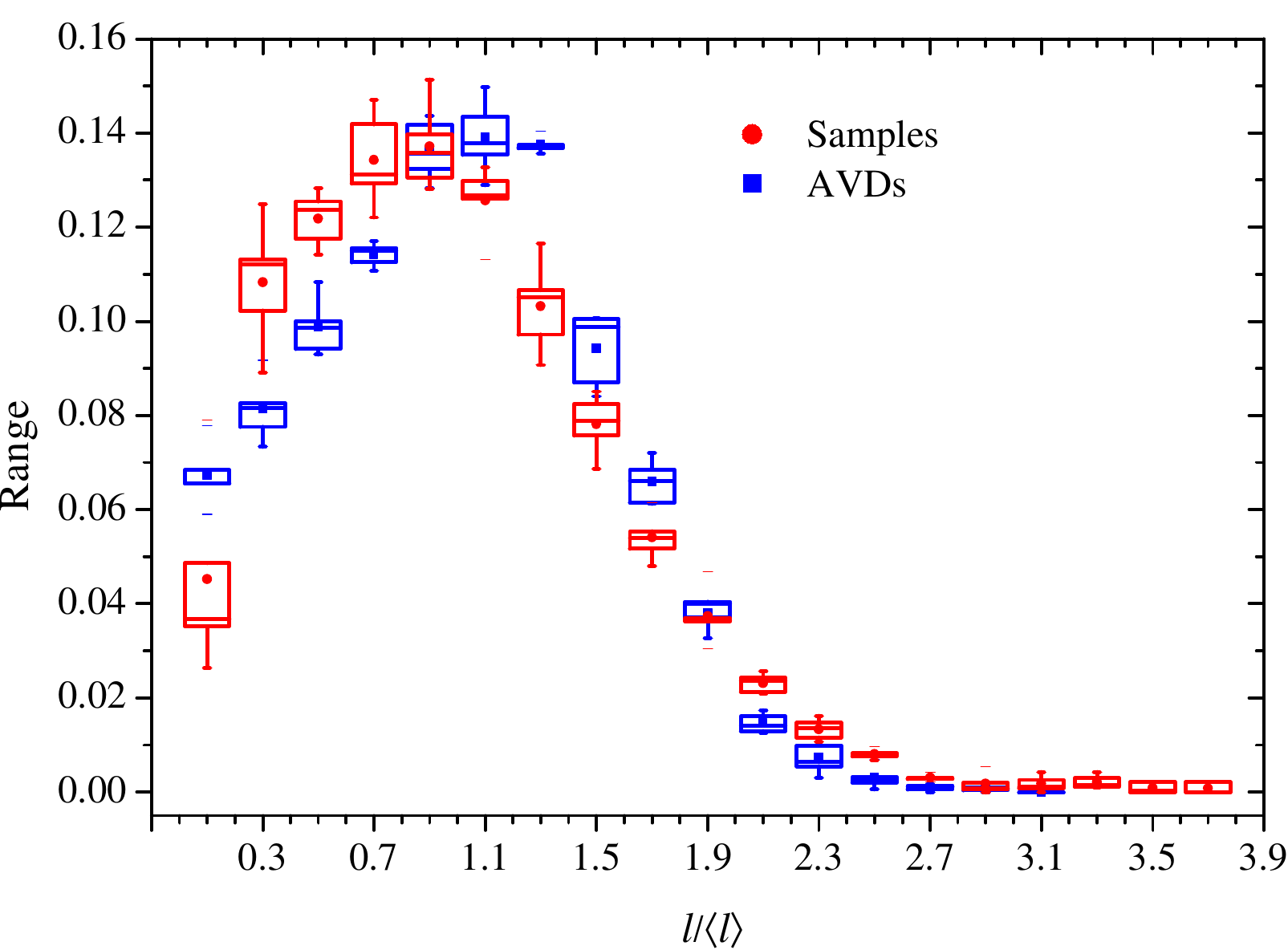}
  \caption{Edge length distributions in samples of real-world crack patterns and in AVDs.}\label{fig:PDFSamples}
\end{figure}

Figure~\ref{fig:PDFSamples} presents the edge length distributions in samples of real-world crack patterns and in AVDs. Not surprisingly, the variance for the case of AVDs is smaller than that found in the original real-world patterns. as the result, the distribution of the face areas is as well narrower for the AVDs as compared to the real samples~\footnote{See Fig. S3 in Supplemental Material for the distribution of the face areas.}

\begin{figure}[!htb]
  \centering
  \includegraphics[width=0.8\columnwidth]{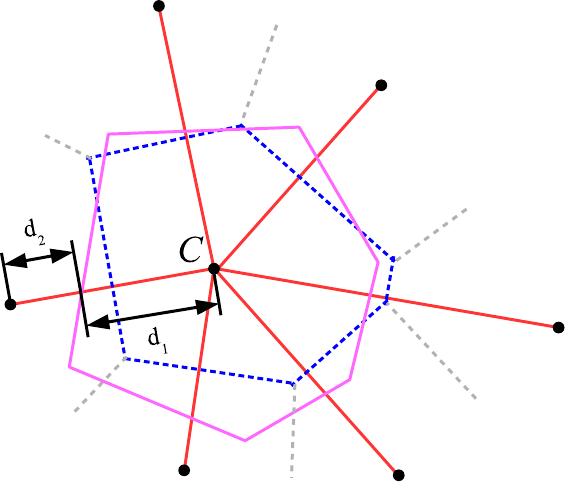}
  \caption{A hexagon drawn using a solid line depicts a face of crack pattern. Point $C$ depicts the centroid of this face, while other points correspond to centroids of neighboring faces. A hexagon drawn using a dashed line depicts a face of Voronoi diagram.}\label{fig:NVN}
\end{figure}

Figure~\ref{fig:NVN} sketches a fragment of AVD, viz., a seed $C$ (centroid) and several other seeds in its nearest neighborhood. The perpendicular bisectors (dashed lines) of line segments connecting this seed $C$ to its neighbors form a cell of Voronoi diagram. To characterize a deviation of this AVD face from the face of the crack network (a hexagon drawn in solid lines), we found on which parts $d_1$ and $d_2$ each segment connecting $C$ and its neighboring centroid is divided by an edge of the crack face.

Obviously, $d_1/d_2 =1$ for any segment in the case of a Voronoi diagram. The quantity
\begin{equation}\label{eq:NVN}
  Z_j = \frac{1}{N_i}\sum_{i=1}^{N_i}\frac{d_{i1}-d_{i2}}{d_{i1}+d_{i2}}
\end{equation}
indicates a deviation of a $j$-th face of AVD from the corresponding face of the crack pattern. Here, $N_j$ is the number of edges in the $j$-th AVD face, the summation goes over all neighboring seeds. In such a way, we got a quantity which characterizes a difference of a crack face on the face of AVD (Tab.~\ref{tab:NVN}). Since the mean value in all cases is close to zero, standard deviation, skewness, and kurtosis  supply us with additional information~\footnote{See Fig. S4 in Supplemental Material for distributions of $Z$ for each sample.}.
\begin{table}[!htb]
  \centering
  \caption{Deviation between crack patterns and their AVDs.\label{tab:NVN}}
\begin{ruledtabular}
\begin{tabular}{lcccc}
 & $\langle Z \rangle$ & $\sigma_Z$ & Skewness & Kurtosis \\
   \hline
sample 1 & -0.03 & 0.18 & -0.64 & 1.0        \\
sample 2 & -0.04 & 0.21 & -0.82 & 1.1        \\
sample 3 & -0.06 & 0.26 & -0.53 & -0.16      \\
sample 4 & -0.04 & 0.22 & -0.50 & 0.25       \\
sample 5 & -0.06 & 0.26 & -0.63 & -0.067     \\
\end{tabular}
\end{ruledtabular}
\end{table}

The analysis we have performed allows us to draw the following conclusions:
\begin{enumerate}
  \item The distribution of centroids does not follow the Poisson distribution, i.e., PVDs are not the appropriate approach to mimic crack patterns.
  \item The centroid locations do not exhibit any long-range order (the structure factor shows only noise, the RDF has a single maximum at short distances), hence, Voronoi diagrams produced using a distorted regular arrangement of seeds~\cite{Priolo1992,Zhu2001,Qiang2024} also are not the best choice to mimic our particular crack patterns.
  \item The centroid locations do not demonstrate hyperuniformity (the variance of the number of centroids inside a circle varies quadratically with the radius of the circle). Thus, Voronoi diagrams generated using hyperuniformly distributed seeds are also not the appropriate choice to mimic crack patterns under consideration.
  \item There are clear restrictions on the centroid locations: the distance between the closest centroids resembles a truncated Gaussian distribution, in particular, there exists a minimum distance closer than which centroids cannot be located.
\end{enumerate}

\subsection{Phenomenological model}
\begin{figure*}[!htbp]
  \centering
  \includegraphics[width=\textwidth]{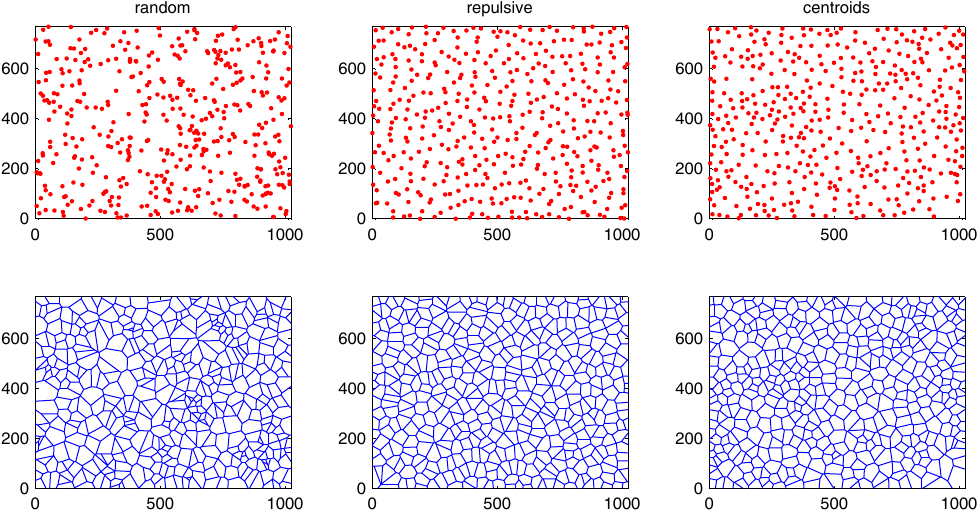}
  \caption{The points are randomly placed (left), while the points are placed with a given probability (center); a set of centroids obtained by processing the real crack pattern (right). The bottom row shows the AVDs.}\label{fig:springs}
\end{figure*}
Let there be a rectangular region on a plane whose linear dimensions are equal to $L_x \times L_y$. We will sequentially and randomly place points in this region, assuming that there is some repulsive force between the already placed points and the newly placed one. We will assume that the maximum interaction distance of the points is equal to $d$. If the minimum distance between the newly placed point and all the already placed points is greater than or equal to $d$, then the attempt to place a new point is considered as successful. If the distances $r_i$ from the newly placed point to $M$ already placed points are less than $d$, then the attempt to place the point is accepted with a probability that depends on both the number of points and the distance to these points. Thus,
\begin{equation}\label{eq:Pspring}
P = \begin{cases}
1, & \text{if} \quad M = 0,\\
\\
d^{-M}\prod\limits_{i=1}^{M} r_i & \text{if} \quad M > 0.
\end{cases}
\end{equation}
On the one hand, the proposed rule is close to the model of `particles with soft shells'. On the other hand, the nature of particle repulsion resembles elastic forces.

Figure~\ref{fig:springs} shows situations when the points are randomly placed in the region (on the left), when the points are placed using the acceptance criterion~\eqref{eq:Pspring} with $d=50$ (in the center); and, for comparison, a set of centroids obtained by processing the real structure of cracks is shown on the right. Upon simulation, the number of seeds was equal to the number of centroids. It is evident that the central and right images are qualitatively similar in appearance (from visual inspection).
%========== fig. 9

\begin{figure}[!htbp]
  \centering
  \includegraphics[width=\columnwidth]{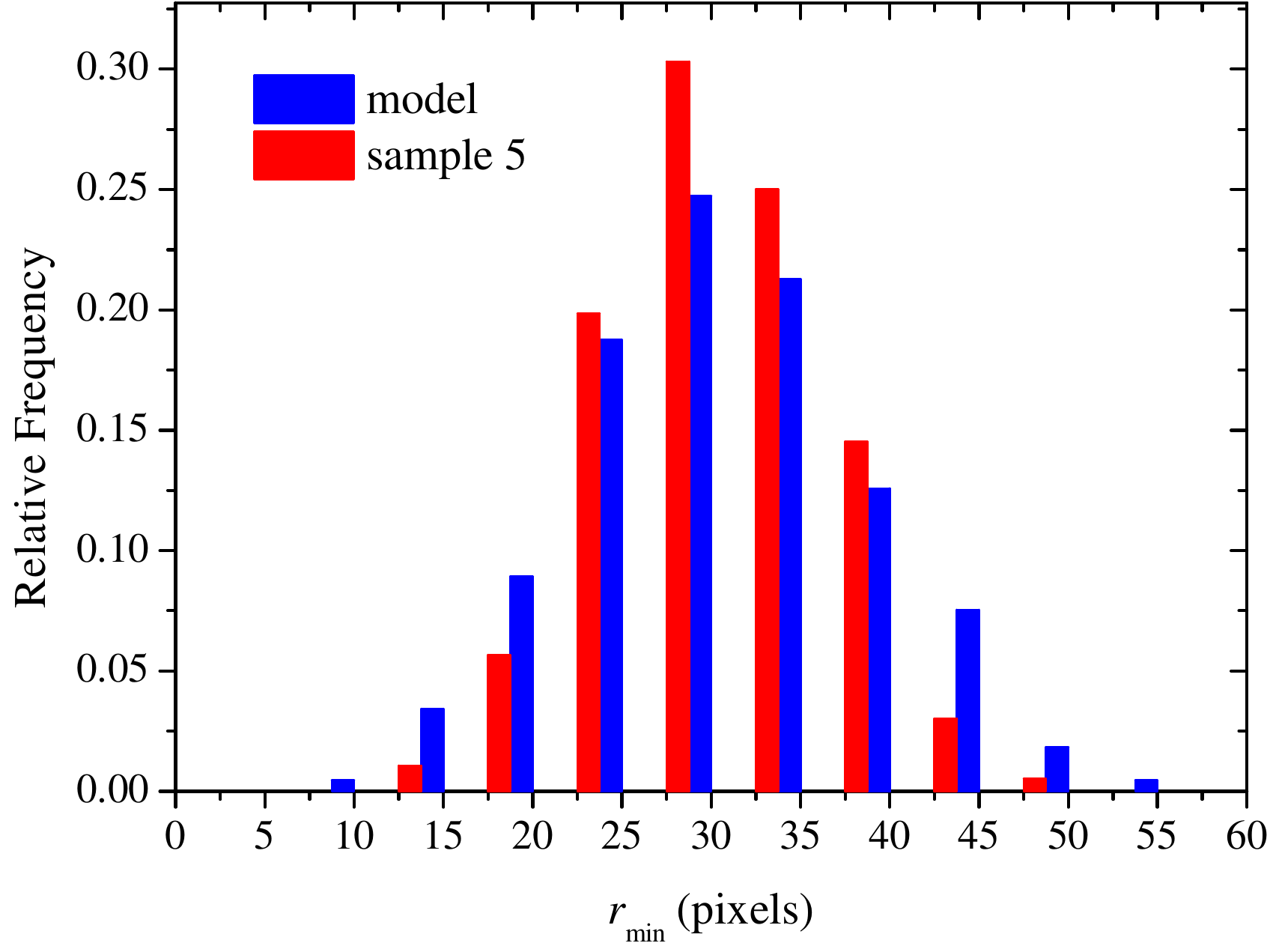}
  \caption{Distribution of distance between nearest points in one particular sample and in a corresponding computer-generated network based on the rule~\eqref{eq:Pspring}, $d=50$.}\label{fig:distribdmin}
\end{figure}

Quantitative aspects of the pattern generated using acceptance criterion~\eqref{eq:Pspring} for the placement of the points are presented in Fig.~\ref{fig:distribdmin} and Tab.~\ref{tab:circularuty}. The distribution of the distance between the nearest points resembles a Gaussian distribution, as was the case for the centroids of networks obtained from the real crack patterns.

\begin{figure}[!htbp]
  \centering
  \includegraphics[width=\columnwidth]{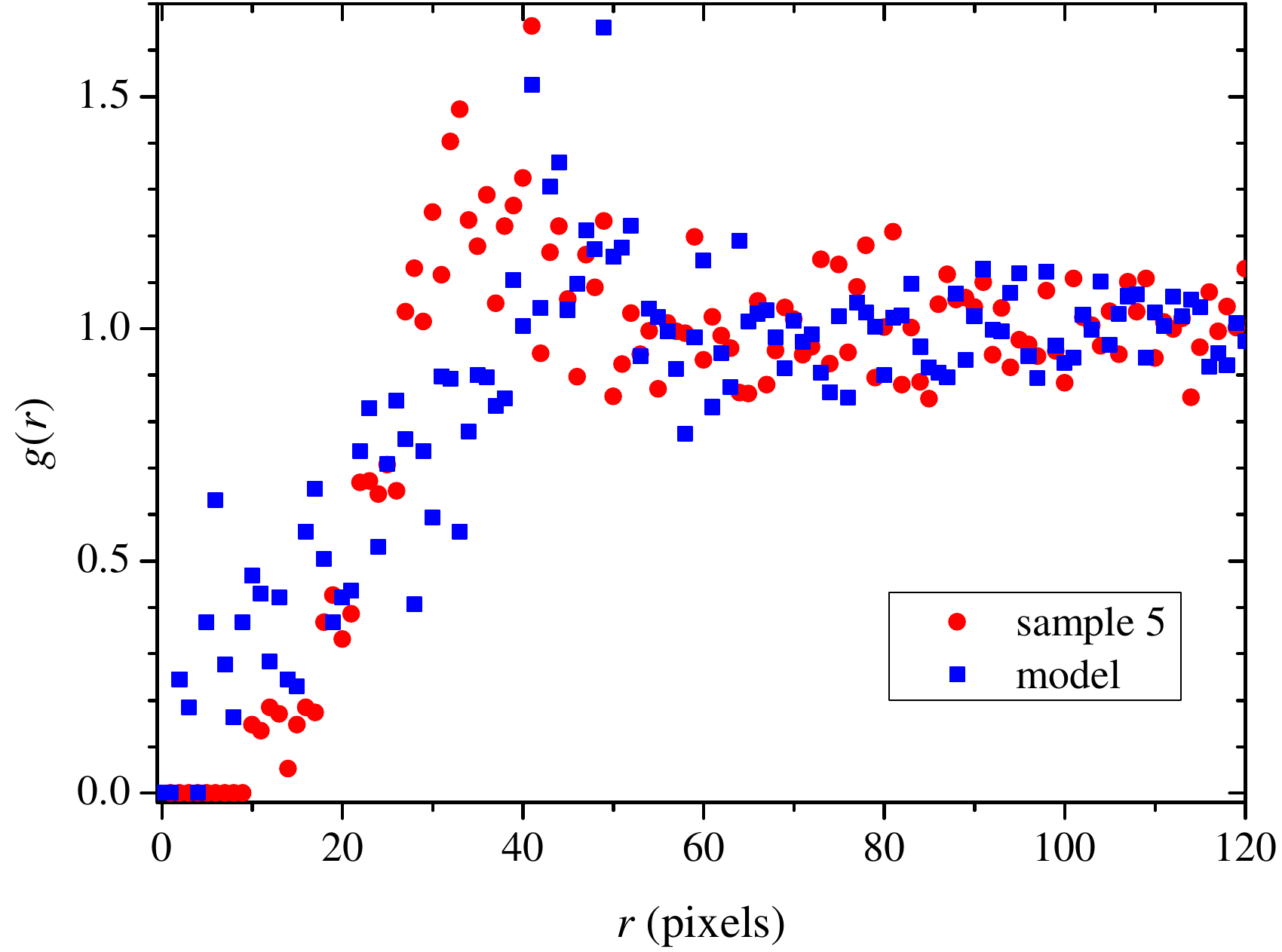}
  \caption{RDF of one particular sample along with RDF of the computer-generated network based on the rule~\eqref{eq:Pspring}, $d=50$.}\label{fig:RDFmodel}
\end{figure}

The RDF presented in Fig.~\ref{fig:RDFmodel}  shows a peak at small distances with no evidence of long range order. This feature resembles closely the behavior seen in the real-world crack patterns.

Unfortunately, the correct placement of centroids does not guarantee the correct distribution of the geometric properties of the network with respect to the real-world crack patterns. In the Voronoi diagram constructed in this way, the main, already noted above, drawback remains: if in real crack networks there is a large variety of cell shapes and sizes and, as a consequence, a large dispersion of the distribution of edge lengths, then in the Voronoi network the cells are of similar sizes and shapes~\footnote{See Fig. S5 in Supplemental Material for edge length distribution.}. To overcome this drawback, it is possible to use a Voronoi tessellation with weights, when different weights are assigned to the seeds, and the edge is drawn not through the midpoint of the line connecting the seeds, but closer to the seed with the greater weight.

\subsection{Electrical conductivity}

\begin{figure}[!htb]
  \centering
  \includegraphics[width=\columnwidth]{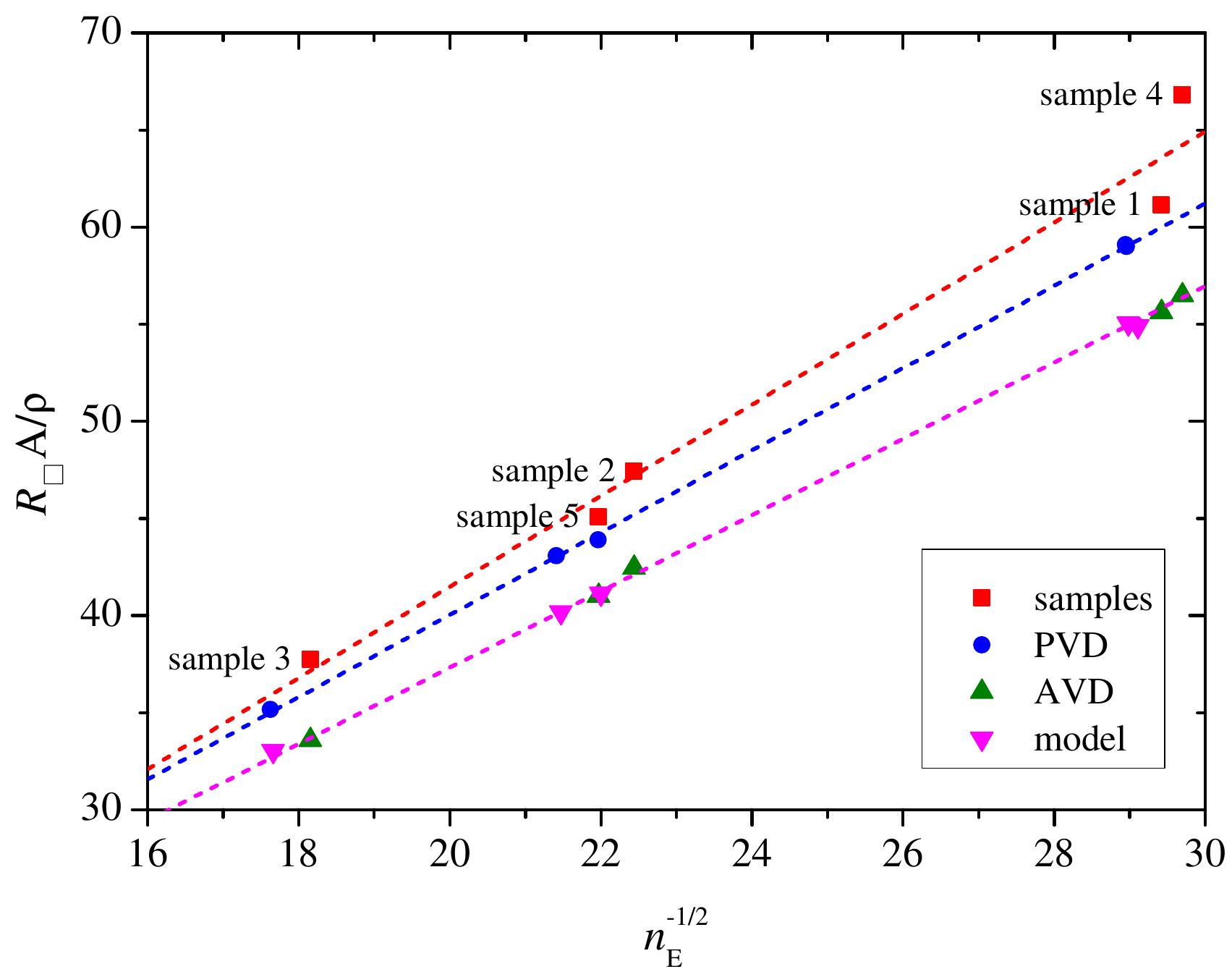}
  \caption{Electrical conductance of RRNs vs the number density of edges.}\label{fig:Rsq}
\end{figure}

Figure~\ref{fig:Rsq} compares the results of the direct computations of the electrical conductance of the (i) crack pattern networks (squares), (ii) corresponding AVDs (triangles), (iii) networks obtained using our model (inverted triangles), (iv) PVDs (circles). Figure~\ref{fig:Rsq} reveals that PVDs fairly well reproduce the electrical resistance of the real crack-based-templates, while  networks obtained using our model have electric resistances that are very close to those of the AVDs. Lines correspond to the least squares fits. Slopes are $1.97 \pm 0.03$ for model and AVD, $2.12 \pm 0.2$ for PVD, and $2.3 \pm 0.2$ for samples.

%=======================================================
The electrical conductivity can be also estimated using an EMT. For our particular purpose, equation \eqref{eq:MarchantEMT} is more convenient to write using resistances rather than conductances~\cite{Nagaya1998}
\begin{equation}\label{eq:MarchantEMTr}
  \sum_i \frac{r_m - r_i}{r_m + r_i (z/2 -1)} = 0.
\end{equation}
Voronoi diagram is a 3-regular network, hence, $z=3$. Since the electrical resistance of a wire is proportional to its length, the PDF of branch resistances corresponds to the edge length PDF of PVD, $f_L(l;n_s)$, where $n_s$ is the number density of seeds.
\begin{equation}\label{eq:gmVoronoi}
\int_0^{l_\text{max}} f_L(l;n_s) \frac{l_m - l }{2 l_m + l } \, \mathrm{d}l = 0.
\end{equation}
The PDF of the edge lengths is known for PVDs in quadratures~\cite{Muche1996,Brakke2005}, the PDF for the unit seed density was calculated~\cite{Brakke2005}. A numerical solution of~\eqref{eq:gmVoronoi} gives $l_m \sqrt{n_s} =0.56963$, thus, $g_m = 1.756 \sqrt{n_s} A/\rho$.

\begin{figure}[!htb]
  \centering
  \includegraphics[width=\columnwidth]{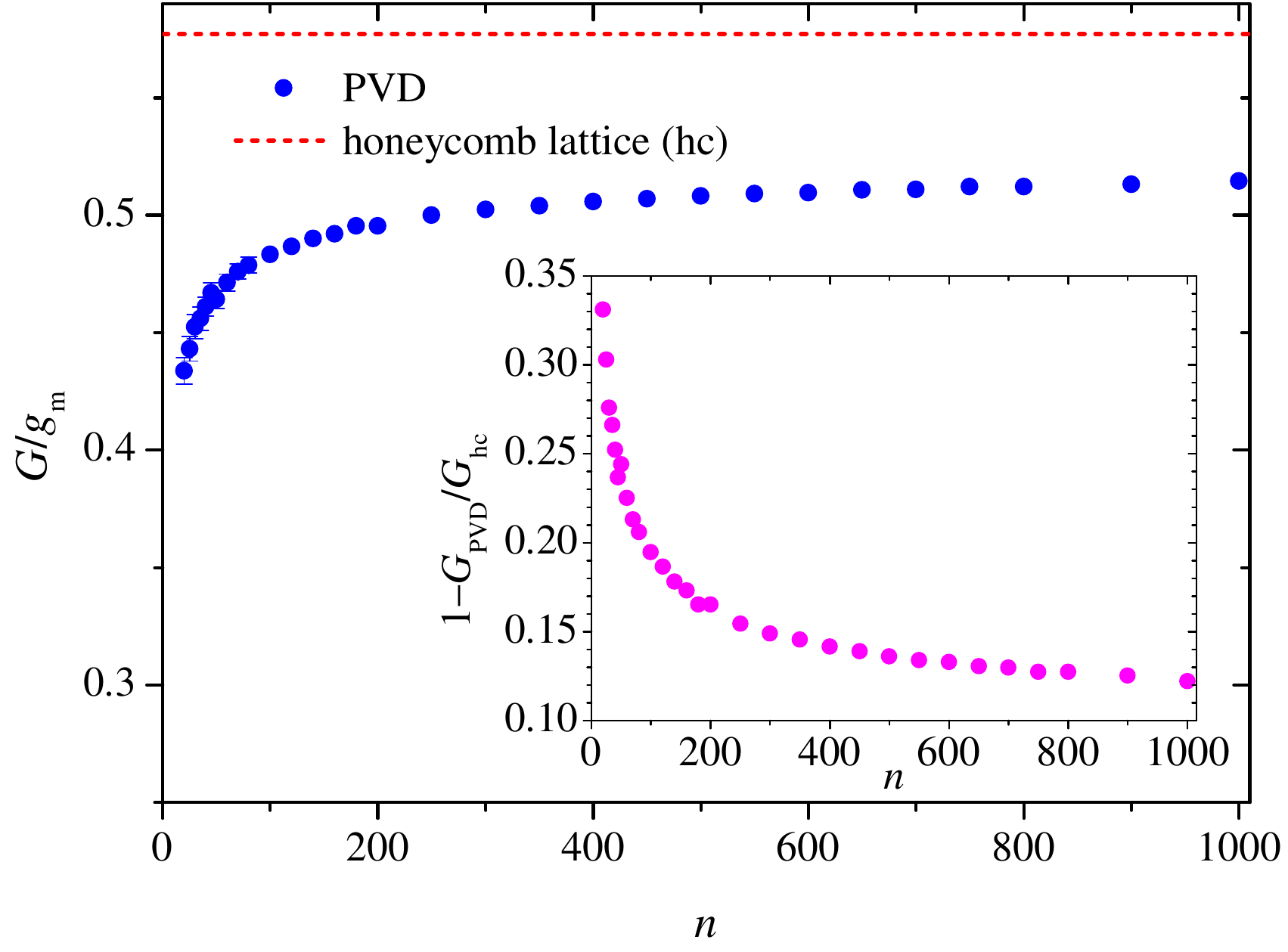}
  \caption{Electrical conductance of RRNs having the structure of PVDs vs the number density of seeds, $n$. Each resistor in the network has the same conductance $g_m$. The results were averaged over 100 independent runs. Electrical conductance of the honeycomb lattice $G_\text{hc}$ (solid line) is presented for comparison. Inset: convergence of the electrical conductance  of RRNs having the structure of PVDs to the value $0.88G_\text{hc}$ as the number density of seeds increases.}\label{fig:GVoronoi1}
\end{figure}

Computations of the electrical conductance for RRNs having the structure of PVD, while all branches have equal values of the resistor, $g_m$, are presented in Fig.~\ref{fig:GVoronoi1}.
%=======================================================

The electrical conductance of the RRNs with equal values of the resistor approaches a constant value as the number density of seeds increases (in other words, when the finite size effect decreases). For dense networks, this value is about 12\% below the electrical conductance of the honeycomb lattice $G_\text{hc} = g_m/\sqrt{3}$. Since, to the best of our knowledge, the analytical dependency of the electrical conductance on the number density of seeds is unknown for the RRNs having the structure of PVDs, the honeycomb lattice may be used for an upper estimate of the electrical conductance of the RRNs under consideration. Hence, the electrical conductance of RRNs having the structure of PVDs can be estimated as $G_\text{PVD} \approx  0.508g_m$.
Thus, given that for a 3-regular pattern with many edges, faces, and vertices, the number densities of seeds and of edges are approximately related as $n_s = n_E/3$, we find
\begin{equation}\label{eq:GPVD}
R_\text{PVD}\approx 1.94 \frac{\rho}{A \sqrt{n_E}},
\end{equation}
which is fairly close to the results of direct computations.

\section{Conclusion\label{sec:concl}}
We studied several images of crack-template-based transparent conductive films and found that the distribution of the centroids of the crack faces centroids resemble distributions arriving upon deposition of repulsive particles rather than deposition of noninteracting particles or hard core particles. The centroids of the crack faces were used as seeds to generate accompanying Voronoi diagrams that are neither Poisson--Voronoi nor centroidal Voronoi diagrams. It bears noting that Poisson--Voronoi and centroidal Voronoi diagrams have both been used as tools for describing crack patterns~\cite{Haque2023}. Our present findings together with published results~\cite{Haque2023} indicate that the variety of morphologies observed in real-world crack patterns in different materials and diverse situations cannot be accurately described within a single, unique, modeling framework. On the contrary, an accurate description of the patterns of cracks formed in materials of different natures is likely to require distinct and individualized models that are tailored to each class of system. As examples, centroidal Voronoi diagrams adequately describe columnar rock patterns~\cite{Li2025}, while STIT~\cite{Leon2020}, Gilbert tessellation, or the cell division algorithm are appropriate for modeling crack patterns in brittle materials~\cite{Haque2023}.

There are significant differences between crack-template-based electrodes and columnar rock patterns and soil crack patterns. Columnar rock patterns and soil/mud/clay desiccation crack are three dimensional systems, in which a material property (temperature in the case of melted rocks/lava or humidity in the case of soil) continuously changes in the vertical direction. By contrast, in the case of crack-template-based electrodes, the systems are quasi two dimensional, while material properties have a jump at the interface between a film and a substrate. Boundary conditions, surface tension, and capillary and interfacial flow have been shown to play an important role in the physics of desiccating cracks and droplets, and can affect the resulting elastic properties. The shape of the domain can thereby influence the patterns that are formed by controlling where cracks are initiated, and droplets and films are different in this respect~\cite{Mondal2023,Parmar2024}.

We proposed a model that allows generating networks having structure similar to desiccation crack patterns which occur during production of crack-template based conductive films. However, since the morphology of such patterns is sensitive to particulars of the technology, materials, and desiccation history, it is conceivable that quantitatively accurate modeling of the variety of patterns that can arise may be beyond the range of a single model to capture. Such details, for instance, might include hierarchical structure of patterns and `brickwork' domains. The problem is that a Voronoi tessellation divides the whole domain at once, while the real cracks occur one by one in a temporal sequence~\cite{Bohn2005,Kumar2021}. Cracks belonging to different generations may have different widths and grow at different rates; widths of the conductive channels (cracks filled with a metal) have direct impact on the sheet resistance of the TCFs. However, an iterative Voronoi tessellation can be easily realised~\cite{tarasevich2025algorithm} in the same manner as an iterative cell division algorithm~\cite{Haque2023} and could even be initialized with a non-Poisson distribution of seeds such as proposed in the present work~\eqref{eq:Pspring}.

We calculated the electrical conductance of the resistor networks, whose structure corresponds to plane Poisson--Voronoi diagrams, while each diagram edge is considered as a conductor having a conductance $g_m$. As the boundary effect decreases (the number density of edges increases), the conductance tends to the value $\approx 0.88 g_m/\sqrt{3}$ (note that $g_m/\sqrt{3}$ is the electrical conductance of an infinite honeycomb lattice). A comparison with the computations of the electrical conductance of crack-template-based networks suggest that Poisson--Voronoi diagrams are reasonable models to mimic electrical properties of such networks.

In order to estimate how the conductance of random resistor networks based on Poisson--Voronoi diagrams depends upon key physical variables, we utilised the effective medium theory in the same manner as~\cite{Marchant1979}. The effective medium theory provides a nice approximation of the dependency of the electrical conductivity on the number density of edges. However, this approximation requires an additional estimate, since, for a network whose structure corresponds to a Voronoi network while all branch resistances are equal, an analytical expression of electrical conductance is unknown.

We show that the geometrical properties are not simply captured by the PVD, and there are features that suggest an apparent repulsion between the centers of the emerging faces. Additionally, the relationship between crack network geometry and sheet conductance is shown to be complicated, given that the PVD provides a slightly better model for the conductivity determined from simulations. Improving the description of certain aspects of the network geometry may not, therefore, by itself lead to a more accurate model for the conductivity.

\begin{acknowledgments}
We acknowledge funding from the Russian Science Foundation, Grant No. 23-21-00074 (I.V.V. and A.V.E.).
\end{acknowledgments}

\bibliography{EMTarticle}

\end{document}